\documentclass[10pt,conference]{IEEEtran}
\usepackage[latin9]{inputenc}
\usepackage{color}
\usepackage{amsmath}
\usepackage{amssymb}
\usepackage{graphicx}

\makeatletter
\IEEEoverridecommandlockouts
\usepackage{cite}
\usepackage{amsfonts}\usepackage{algorithmic}
\usepackage{textcomp}
\usepackage{xcolor}
\begin{document}
\title{An Open Platform for Simulating the Physical Layer
of 6G Communication Systems with Multiple Intelligent Surfaces}
\author{\IEEEauthorblockN{Alexandros Papadopoulos} \IEEEauthorblockA{\textit{CSE Dept.} \& Information Technologies Institute \\
 \textit{University of Ioannina} \& CERTH\\
 Ioannina, Greece \\
 a.papadopoulos@uoi.gr} \and \IEEEauthorblockN{Antonios Lalas} \IEEEauthorblockA{\textit{Information Technologies Institute}\\
  \textit{CERTH}\\
 Thessaloniki, Greece \\
 lalas@iti.gr} \and \IEEEauthorblockN{Konstantinos Votis} \IEEEauthorblockA{\textit{Information Technologies Institute}\\
  \textit{CERTH}\\
 Thessaloniki, Greece\\     
 kvotis@iti.gr} \and \IEEEauthorblockN{Dimitrios Tyrovolas} \IEEEauthorblockA{\textit{ECE Dept.} \\
 \textit{Aristotle University of Thessaloniki}\\
 Thessaloniki, Greece\\ 
 tyrovolas@auth.gr}\and \IEEEauthorblockN{George Karagiannidis} \IEEEauthorblockA{\textit{ECE Dept.} \\
 \textit{Aristotle University of Thessaloniki}\\
 Thessaloniki, Greece\\
 geokarag@auth.gr}\and \IEEEauthorblockN{Sotiris Ioannidis} \IEEEauthorblockA{\textit{ECE Dept.} \\
 TUC \& FORTH\\
 Chania, Greece\\
 sotiris@ece.tuc.gr}\and \IEEEauthorblockN{Christos Liaskos} \IEEEauthorblockA{\textit{CSE Dept.} \\
 \textit{University of Ioannina} \& FORTH\\
 Ioannina, Greece \\
 cliaskos@uoi.gr} }

\maketitle
\begin{abstract}
Reconfigurable Intelligent Surfaces (RIS) constitute
a promising technology that could fulfill the extreme performance
and capacity needs of the upcoming 6G wireless networks, by offering
software-defined control over wireless propagation phenomena. Despite
the existence of many theoretical models describing various aspects
of RIS from the signal processing perspective (e.g., channel fading
models), there is no open platform to simulate and study their actual
physical-layer behavior, especially in the multi-RIS case. In this
paper, we develop an open simulation platform, aimed at modeling the
physical-layer electromagnetic coupling and propagation between RIS
pairs. We present the platform by initially designing a basic unit
cell, and then proceeding to progressively model and simulate multiple
and larger RISs. The platform can be used for producing verifiable
stochastic models for wireless communication in multi-RIS deployments, such as vehicle-to-everything (V2X) communications in autonomous vehicles and cybersecurity schemes, while its code is freely available to the public. 
\end{abstract}


\section{Introduction}

The coming of 6G wireless communications renders as priority the
realization of the, until now, uncontrollable communication environment
as a programmable resource~\cite{CACM.2018}. The utilization of
the metasurfaces--recently popularized in the communications society
under the term Reconfigurable Intelligent Surfaces (RISs)~\cite{di2020smart}--could
play a game changer role in the manipulation of the communication
field. The key indicator is that, instead of considering environmental
reflection and scattering as uncontrollable phenomena whose impacts
can only be described stochastically, they could be considered as
parameters of the network that should be optimized~\cite{Liaskos.TNET.2019}.

RISs mainly are planar and rectangular, tile-resembling
devices whose physical characteristics, and particularly their permittivity
and permeability, can be engineered in real-time to obtain a required
macroscopic electromagnetic behavior~\cite{Liaskos.TNET.2019}. Thus,
when a wireless wave emitted by a user device (e.g., a smartphone)
impinges upon an RIS, it can be programmatically reflected to any
custom direction, or even be split among several ones. Based on this
property, a popular demo case of the RISes is the establishment of
\emph{effective LOS (line of sight) conditions}
between a transmitter and a receiver who are actually in NLOS~\cite{Liaskos.TNET.2019}.
Furthermore RIS uses include the mitigation of multi-path fading phenomena,
the overcoming of localized coverage holes and the optimal energy
management in IoT systems~\cite{Liaskos.TNET.2019,basar2019wireless}.
Moreover, RISs could assist in the design of low-complexity and energy-efficient
massive Multiple-Input-Multiple-Output (MIMO) transmitting and receiving
antennas, meeting a very promising expectation of 5G and Beyond-5G
(B5G) networks~\cite{wymeersch2020beyond}. Since RISes can act as
generic facilitators of the communication links between base stations
and the end-users, their system use cases are abundant. Smart homes,
cities, hospitals and industries are theorized to benefit greatly
via the RIS technology. Moreover, the utilization of RISes in vehicle-to-everything
(V2X) communications could overcome the current challenges stemming
from the acute fading and Doppler shift suffered is these systems~\cite{b9,b4}. 

Due to their significant potential, several RIS-enabled
channel models have been proposed. Indicatively: in \cite{b2}, multiple
RIS are employed to support Non-Orthogonal-Multiple-Access (NOMA)
networks; in \cite{b3}, the closed-form expressions for the behavior
of cascaded multi-RISs are presented; in \cite{b8}, the performance
of a single RIS for time-division multiple access (TDMA), frequency-division
multiple access (FDMA), and NOMA is presented. However, despite multiple
channel models proposed, simulating the physics of such a system remains
a task for the physicist, and not usable by wireless engineers in
general. 

The contribution of this paper is a physics simulation
platform for studying the communication between any RIS couple of
a multi-RIS deployment. The platform is open and freely available~\cite{b13},
and provides the tool for exploring the wireless propagation environment
between RIS units in any ecosystem. This can be used for deducing
accurate wireless channel models, validated via actual physics simulations, enabling a variety of application domains, such as V2X communications, especially in autonomous vehicles (AVs), as well as cybersecurity defence mechanisms. 

The remainder of this paper is organized as follows.
In Section~\ref{sec:System-model-and} we present the RIS system
model employed by the platform. In Section~\ref{sec:Platform-Design-Methodology}
we present the design methodology of the proposed platform. Section~\ref{sec:Evaluation}
demonstrates use cases built with the platform, and the paper is
concluded in Section~\ref{sec:Conclusion-and-Future}.

\section{System Model and Related Studies}\label{sec:System-model-and}

The proposed platform models the accurate electromagnetic
propagation between two RIS units--whose composition, dimensions,
and state are user-defined--while varying the distance between them.
Each RIS consists of a planar arrangement of elements, 
known as cells in the metasurface terminology~\cite{MSSurvey}. In
the following, we employ the Square Split Ring Resonators~\cite{milias2021metamaterial} (S-SRR)
design for each RIS cell as a running example and for ease of elucidation.
Each element hosts a port and a lumped element offering tunable impedance
within a user-specified range. We consider that all the elements of
the RIS$_{1}$ are active, meaning that power is generated at each
of their ports, enters the system and is altered by the state of the
corresponding lumped element. Additionally, all the elements of the
RIS$_{2}$ are passive, meaning that their ports receive power emanated
from RIS$_{1}$, alter it based on the state of the local lumped element
and re-emit it. Depending on its configuration, this setup emulates
either: i) the RIS-RIS part of a transmitter-RIS-RIS-receiver communication,
which is useful for deriving verifiable stochastic models for the
inter-RIS part of a channel. ii) the transmitter-RIS or the RIS-receiver
part of the communication. (E.g., in order to study the transmitter-RIS
part, one simply sets the number of elements of RIS$_{1}$ to 1). 

Accounting for the capability to freely vary the
distance between the two RIS units (center-to-center), the proposed
platform becomes a versatile tool for uniformly studying any part
of a multi-RIS system. In any customization, the platform treats the
element lumped states as optimization parameters and seeks to maximize
the power transfer from RIS$_{1}$ to RIS$_{2}$ for the given physical
RIS composition and distance. The outputs of the simulation are the
s-parameters from every port of RIS$_{1}$ to every port of RIS$_{2}$.
The user can then define any post-processing of the produced s-parameters,
e.g., for producing an inter-RIS path-loss model versus their distance.
At the core of the platform, we utilize the open-source Finite-difference time-domain (FDTD) solver
OpenEMS~\cite{b5}. 

In our past work, we developed an experimental testbed
for evaluating networked RIS units, without, however, including a
multi-RIS simulation platform~\cite{liaskos2020internet}. On the
other hand, there is an extensive body work dedicated to the stochastic
modeling of RIS-enabled channels that would benefit from such a tool.
Indicatively, in \cite{b6}, the programming of the reflection coefficients
of the unit cells is examined for the optimal performance of the metasurfaces.
In \cite{b7}, indoor and outdoor scenarios are examined in various
frequency bands. In \cite{b8}, the performance of a single RIS for
time-division multiple access (TDMA), frequency-division multiple
access (FDMA), and NOMA is presented. In \cite{b9}, theoretical models
for the optimization and resource allocation in V2X networks are examined.
Lastly, SimRIS Channel Simulator is presented in \cite{b11}. This
tool generates the matrices of channels coefficients among the transmitter,
a single RIS and receiver assuming theoretical channel models as its
basis (e.g., between elements), and without actually simulating the
underlying physics. To the best of our knowledge, there is no structured
and open platform for studying the physical layer of RIS communication.
We make the source code of our platform freely available in Github
\cite{b13}.

\section{Platform Design and Methodology}\label{sec:Platform-Design-Methodology}

\subsection{RIS pair optimization Method}

\begin{figure}[t]
\centering{}\includegraphics[width=6cm,height=3cm]{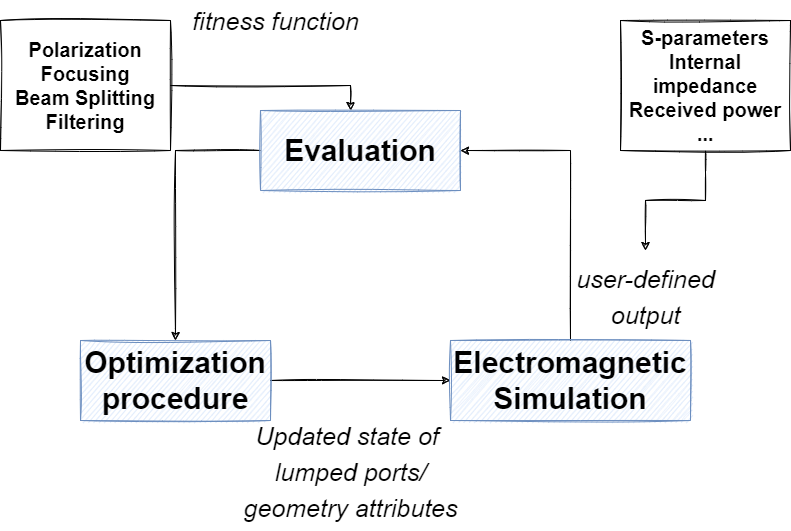}\caption{Computational workflow of optimization procedure.}\label{fig2}
\end{figure}
The general optimization workflow is depicted in Figure~\ref{fig2}. The goal of the optimization procedure is the detection of the optimal configuration of the RIS unit. As configuration, the state of the unit cell's lumped ports or the geometry attributes of the RISs could be considered. The candidate configuration is inserted to the electromagnetic simulation module. In this component, the physical layer of the RIS units is modelled. The user-defined outputs, e.g., s-parameters, internal impedance, received power etc, are utilized for the evaluation of the system's performance. In the evaluation component, the simulator's output is compared  with a given fitness function that describes the desired functionality of the RIS elements. The RIS unit is able to support multiple functionalities such as the change of the reflected wave polarization, the focusing of the reflected waves to a specific direction, the beam splitting in more targets or the filtering of a specific frequency band. \par
The result of the evaluation processing is directed again in the Optimization Procedure module in order the new candidate configuration to be generated. In this step, any optimization method, e.g., a Genetic Algorithm, a Differential Evolution or a Particle Swarm Optimization, could be utilized. The procedure ends once the level of fitness is deemed as satisfactory or the maximum available amount of time is devoted.

\subsection{Electromagnetic Simulation}
The electromagnetic simulation could be established in the proposed simulation platform that consists of two main parts; the \emph{unit cell definition} and the \emph{RIS pairs simulation}. Initially, \emph{unit cell definition} is utilized for the design of the unit cell. The unit cell consists of the substrate, the groundplane and the S-SRR. The properties of the substrate are user-defined while the groundplane and the S-SRR are layers of metal. In the gap of S-SRR's outer ring, a lumped port is positioned. This port could radiate or not. The radiation is a Gaussian excitation. \par 

\begin{figure}[t]
\centering{}\includegraphics[width=5cm,height=5cm]{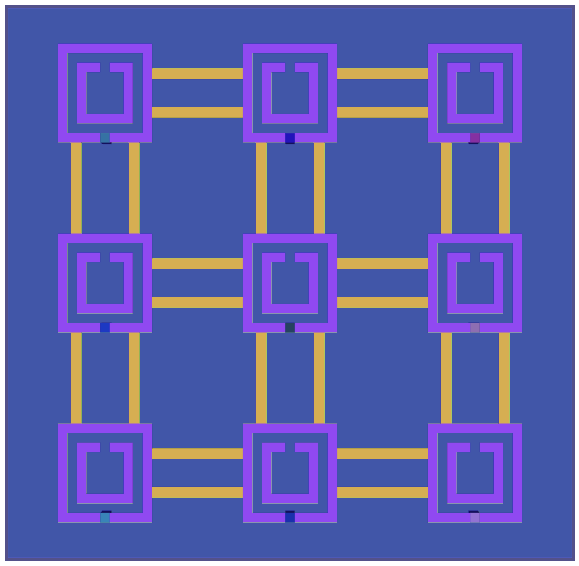}\caption{3x3 RIS with load patches.}\label{fig9}
\end{figure}
In the next step, the \emph{RIS pairs simulation} creates two identical metasurfaces composed by periodically positioned, identical, unit cells. The S-SRRs are connected with the adjacent ones via load patches (Figure~\ref{fig9}). The distance between the RIS pairs is, also, user-defined. \par
The s-parameters is the basic output of the simulator, which can then be post-processed into any fitness function of interest, to guide the optimization as required by the user. Generally, the s-parameters in a network express the transfer of power from an active port (source) to another that could be passive or not (reference). When the port that acts as source is identical with
the reference, the corresponding parameter is called reflection coefficient.
In the case that the source and the reference is considered in different
ports, the respective parameter is called transmission coefficient. The transmission coefficient is a measurement about the energy coupling between an active and a passive port. The simulation calculates the following:
\begin{itemize}
\item The feed point impedance of the active ports.
\item The incoming, reflected and accepted (subtraction of incoming and reflected) power in both active and passive ports.
\item The reflection coefficients of the active ports and the transmission coefficients of them with the passive ones (s-parameters).
\item The resonating frequency in which the reflection coefficients of the active ports are minimized.
\item The values of frequency in which the active ports exhibit the maximized transmission coefficients with the passive ones.
\end{itemize}

\subsection{Unit cell definition}
\begin{figure}[t]
\centering{}\includegraphics[width=3.5cm,height=3.5cm]{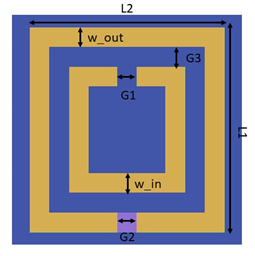}\caption{Unit cell's substrate (blue) and S-SRR (yellow).}\label{fig1}
\end{figure}

The first step is the design of the RISs' unit cell. In the \emph{unit cell definition}, the user is able to re-adjust:
\begin{itemize}
\item The central frequency and the bandwidth of the Gaussian excitation. 
\item The dimensions of S-SRRs (Figure~\ref{fig1}).
\item The properties (electric permittivity, $\epsilon_{r}$, and tangent loss, \emph{tand}) and the dimensions of substrate (width, length and thickness).
\item The distance between the pair of unit cells. By default, the distance is 10 mm.
\end{itemize}
 
The resonating frequency displays strong dependency with the length, the width and the distance of the S-SRRs' rings. The widths of the rings' patch and the gaps do not impact significantly on the behavior of S-SRRs' resonating frequency. As concerns the substrate, its dimensions affect directly and strongly both the resonating frequency and the coupling between the active and the passive ports. Hence, we propose the dedicated variables to be utilized wisely. \par

The fine tuning evaluation is based on s-parameters and the feed point impedance. 
Having defined the port of the first unit cell active and the other one's passive, we expect to observe minimization of the reflection coefficient and maximization of the transmission one at the central frequency. As concerns the feed point impedance, the internal resistance of ports has been set to 50 $\Omega$. Therefore, the real and imaginary part of feed point impedance should be as close as possible to 50 $\Omega$ and 0$ \Omega$, correspondingly, in order the maximum power transfer in the central frequency to be achieved.
\subsection{RIS pairs simulation}
The first action in the \emph{RIS pairs simulation} is the definition of the S-SRRs dimensions as they have resulted from the \emph{unit cell definition} procedure. Thus, the user is able to re-adjust:
\begin{itemize}
\item The dimension of the RIS pairs. The dimension displays the number of the unit cells per row and column.
\item The distance between the RIS units.
\item The distance between the S-SRRs.
\item The width and the thickness of the load patches.
\end{itemize}
The aforementioned parameters can configure the resonating frequency of the active ports and the coupling of them with the passive ones.
 
\subsection{Meshing in OpenEMS}

In the FDTD solvers, a main task is the design of the optimal meshing. The meshing is a strictly rectilinear structure that separates the observed area into many, small cells.
Its optimal design ensures the minimization of the running time and
the maximization of the measurements accuracy. We have adapted the meshing
structure in order the simulation of a multiple-component structure
to be feasible. The main logic is based on the following statements:
\begin{itemize}
\item The resolution expresses the size of the minimal meshing cells. We have introduced two resolution values. The maximum resolution is utilized for the finer structures, e.g., the gaps of S-SRRs. In the coarser structures, e.g., in the surrounding air, we use the coarse resolution value. 
\item We implement a function that intervenes to avoid a long-time simulation in the case that the distance between adjacent meshing lines becomes less than a pre-defined threshold in each axis. By default, this threshold is equal to the half of the maximum resolution.
\end{itemize}

\section{Evaluation: Use Cases}\label{sec:Evaluation}
\begin{figure}[t]
\centering{}\includegraphics[width=6cm,height=3cm]{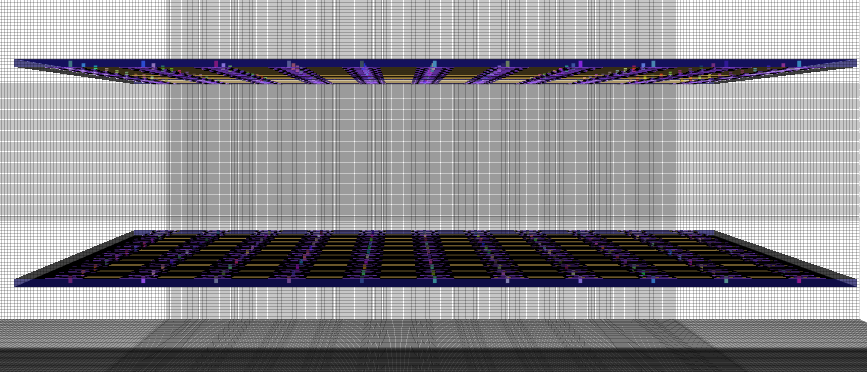}\caption{The format of the 11x11 RIS pairs.}\label{fig11}
\end{figure}

In this section, we use the simulation platform for the modeling of 11x11 RIS pairs (Figure~\ref{fig11}). This setup can exemplary be used for studying two RIS units placed on opposing walls in a floor plan. The central frequency was selected at 8 GHz and the bandwidth at 2 GHz.\par
\begin{figure}[t]
\centering{}\includegraphics[width=7cm,height=5cm]{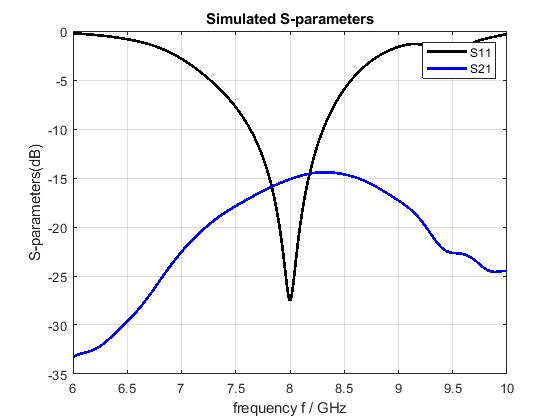}\caption{S-parameters of the S-SRR.}\label{fig5}
\end{figure}
\begin{figure}[t]
\centering{}\includegraphics[width=7cm,height=5cm]{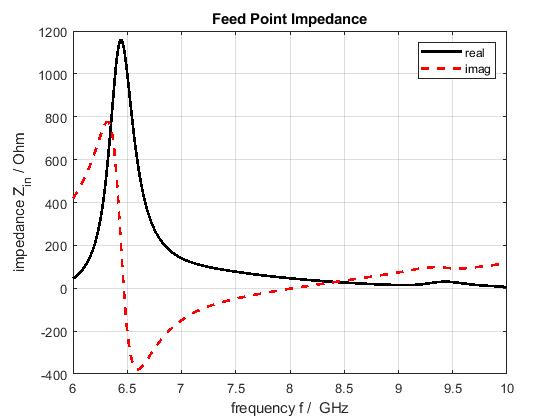}\caption{Feed point impedance of the S-SRR.}\label{fig6}
\end{figure}

\emph{Step 1}: We utilize the \emph{unit cell definition} in order the fine tuning to be achieved. We use a substrate with $\epsilon_{r}$=2.2 and tangent loss tand=0.024.
After the required tests, for S-SRR to resonate at the frequency of 8 GHz, the length of the outer ring must be L2=10.36 mm and the respective width L1=10.9 mm. The gap of the outer and the inner ring is G1=G2=1.05 mm. The width of the S-SRRs patches is $w_{out}$=$w_{in}$=1.05 mm. As concerns the dimension of the substrate, its length is concluded to be equal to its width, at 12.21 mm, and its thickness is  4.8 mm. Observing Figures~\ref{fig5} and~\ref{fig6}, the S-SRRs function properly in the central frequency. This indicates that the fine tuning procedure has been completed.  \par


\emph{Step 2}: Before we proceed to the simulation of the 11x11 RIS pairs, we work on the 3x3 structure in order the fine tuning of the multiple components to be examined in less time and with lower computational resources. We keep the calculated dimensions of the S-SRR stable. \par
Using the \emph{RIS pair simulation}, we concentrate on the dimensions of load patches and the distance between the adjacent S-SRRs. As it has already been mentioned, these variables have great impact on the resonating frequency and the coupling. After the required tests, the width of the load patches is 1.2 mm, very close to S-SRR's patch width. The distance between the S-SRRs in both horizontal and vertical axis is 10 mm. This value is about equal to the half of the wavelength in the substrate material. An extra option that is available to the user in order to reinforce the coupling is the limitation of the substrate's thickness.\par
\begin{figure}[t]
\centering{} \includegraphics[width=7cm]{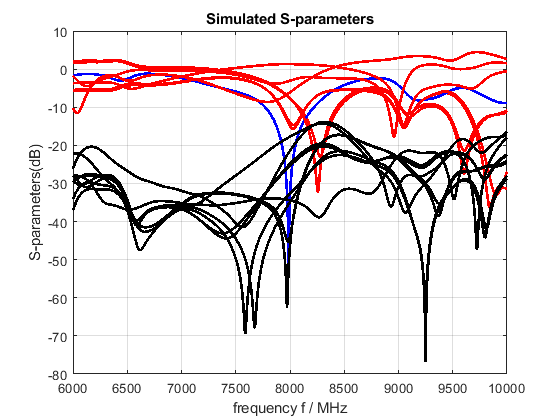}
\caption{S-parameters of an active port with patches in 3x3 RIS pairs.}
\label{fig10}
\end{figure}

In Figure~\ref{fig10}, the blue line represents the reflection coefficient, the red ones depict the transmission parameters among the active elements and the black ones among the passive. It is clear that there is coupling between the active and the passive ports. However, the existence of many active components results in that the resonance is not held in the central frequency for all the unit cells. Despite the fact that there in no common behavior for all the unit cells, it is observed that these ones, which are geometrically symmetric, act similarly (resonating frequency and coupling with the ports). \par

\begin{figure}[t]
\begin{centering}
\includegraphics[width=7cm]{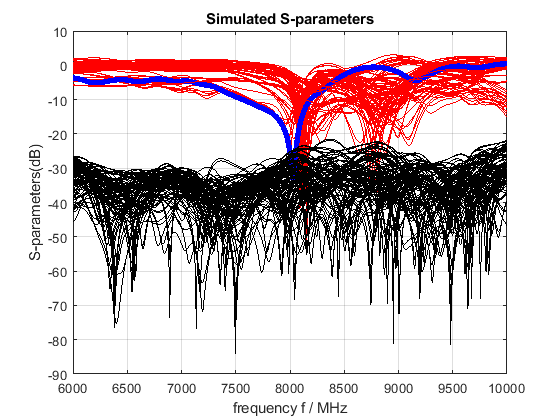} 
\par\end{centering}
\caption{S-parameters of an active port in 11x11 RIS pairs.}\label{fig12}
\end{figure}

\emph{Step 3}: In this phase all the parameters of the system have already been defined. In the \emph{RIS pair simulation}, we just change the dimension of the RIS in order the 11x11 setup to be created. In the Figure~\ref{fig12}, the behavior of an active port is depicted. We can point out that the majority of the active ports exhibit resonances in two bands; one with central frequency at 8 GHz with bandwidth about 200 MHz and another one with central frequency at 8.9 GHz with bandwidth about 250 MHz. \par
Once the design is completed, the behavior of the RIS pairs in different distances could be investigated without any update in the structure. The products of this simulation compose a sufficient dataset for the building of custom fitness functions, and the implementation of the optimization procedure accordingly.

\section{Conclusion and Future Work}\label{sec:Conclusion-and-Future}

In this paper, a simulation platform for the study
of multiple RIS deployments in the physical layer is presented. The
proposed platform is able to simulate RIS pairs of any composition
and dimensions, and at varying distances between them. The platform
provides facilities for pinpointing the resonating frequency of the
defined RIS pair, and then for optimizing the energy flow from one
to the other. The optimization can be customized for any fitness criterion,
and for any tunable state supported by the user-defined RIS designs.
The intended use is the exploitation of the produced simulation data
for validating theoretical channel models of RIS-empowered system
setups. Due to its supported parameterization, the platform can be
employed not only for studying the RIS-RIS communication, but also
for the MIMO(user)-RIS part of a communication. 

Future work is directed towards showcasing the potential
of the platform towards creating realistic channel models for indoor
and outdoor, RIS enabled environments, such as V2X infrastructure in AVs, enabling also cybersecurity aspects.

\section{Acknowledgment}
This work was funded by the projects WISAR (Foundation
for Research and Technology--Synergy Grants 2022) for theoretical design, SHared automation Operating models for Worldwide adoption (SHOW) under Grant Agreement No. 875530, for practical design and applicability study on vehicular communications, and COLLABS (EU Horizon 2020, GA 871518) for applicability study on cybersecurity aspects.

\bibliographystyle{ieeetr}
\bibliography{refs}

\end{document}